# T-MSD: An improved method for ionic diffusion coefficient calculation from molecular dynamics

Yuxiang Gao[1,2,#], Xiaodong Cao[1,2,#], Zhicheng Zhong[1,2,*]

[1] *School of Artificial Intelligence and Data Science, University of Science and Technology of China, Hefei, 230026, China*

[2] *Suzhou Institute for Advanced Research, University of Science and Technology of China, Suzhou, 215123, China*

[*]*Corresponding author. E-mail:* [zczhong@ustc.edu.cn](mailto:zczhong@ustc.edu.cn)

[#]*These authors contributed equally to this work.*

**Abstract**

Ionic conductivity is a critical property of solid ionic conductors, directly influencing the performance of energy storage devices such as batteries. However, accurately calculating ionic conductivity or diffusion coefficient remains challenging due to the complex, dynamic nature of ionic motion, which often yield significant deviations, especially at room temperature. In this study, we propose an improved method, T-MSD, to enhance the accuracy and reliability of diffusion coefficient calculations. Combining time-averaged mean square displacement analysis with block jackknife resampling, this method effectively addresses the impact of rare, anomalous diffusion events and provides robust statistical error estimates from a single simulation. Applied to large-scale deep-potential molecular dynamics simulations, we show that T-MSD eliminates the need for multiple independent simulations while ensuring accurate diffusion coefficient calculations across systems of varying sizes and simulation durations. This approach offers a practical and reliable framework for precise ionic conductivity estimation, advancing the study and design of high-performance solid ionic conductors.

**Introduction**

Solid ionic conductors with high ionic conductivity are crucial for electrochemical energy storage devices, particularly in batteries, as they enable faster charge/discharge rates and greater energy storage capacity [1-11]. However, the immobility of the anionic framework in solid ionic conductors limits ionic motion, which occurs as a rare event involving the hopping of ions between sites with low probability. This dynamic process is highly complex and sensitive to the chemical composition, point defects, crystal structure, and microstructure [12,13]. As a result, accurately predicting the ionic conductivity or diffusion coefficient of solid ionic conductors is particularly challenging.

Previous theoretical works have proved the capability of molecular dynamics (MD) simulations in investigating ionic motion in solid ionic conductors [14-20]. Ionic conductivity is commonly calculated using the Einstein relation by estimating diffusion coefficient from linear regression of mean square displacement (MSD) curves. The accuracy of these calculations is highly dependent on the quality of the MSD data, which ideally should represent the typical diffusive motion of ions over sufficiently long simulation times to capture both normal and anomalous behaviors [21]. Recently, the use of machine-learning force fields for MD simulations has become increasingly popular, as these methods offer first-principles-level accuracy and address the limitations in time- and size-scales, effectively reducing the statistical uncertainty in diffusion coefficient calculation [22-27]. However, several inherent challenges remain in the accurate calculation of diffusion coefficient, such as the achievement of time translation invariance and the treatment of time correlation within trajectories. Therefore, specialized statistical techniques are necessary to reduce the influence of these rare events and ensure accurate estimates of diffusion coefficient.

Here, we introduce an improved post-processing method, T-MSD, designed to address the effects of rare events on correlated data and enhance the statistical reliability in estimating diffusion coefficient. This method comprises two parts, the time-averaged MSD analysis and the block jackknife (BJ) resampling. Using deep potential molecular dynamics (DPMD) simulations, we demonstrate that the time-averaged MSD effectively reduces data fluctuations and achieves time translation invariance, yielding a more robust estimate of diffusion coefficient. To the best of our knowledge, while this method has been utilized to analyze single particle tracking in biology and chemistry fields [28,29], it has rarely, if ever, been applied in the solid-state ionics. Moreover, BJ resampling provides standard error by explicitly accounting for the contributions of correlated trajectories from a single simulation, whereas conventional MSD analysis typically requires multiple independent simulations for the same purpose. Furthermore, the method is robust across systems with varying sizes and MD simulations of different total durations. With this straightforward strategy, one can easily incorporate proper error estimates into the calculation of diffusion coefficient from MD simulations.

## Results and discussions

### Conventional MSD calculation and *D* estimation

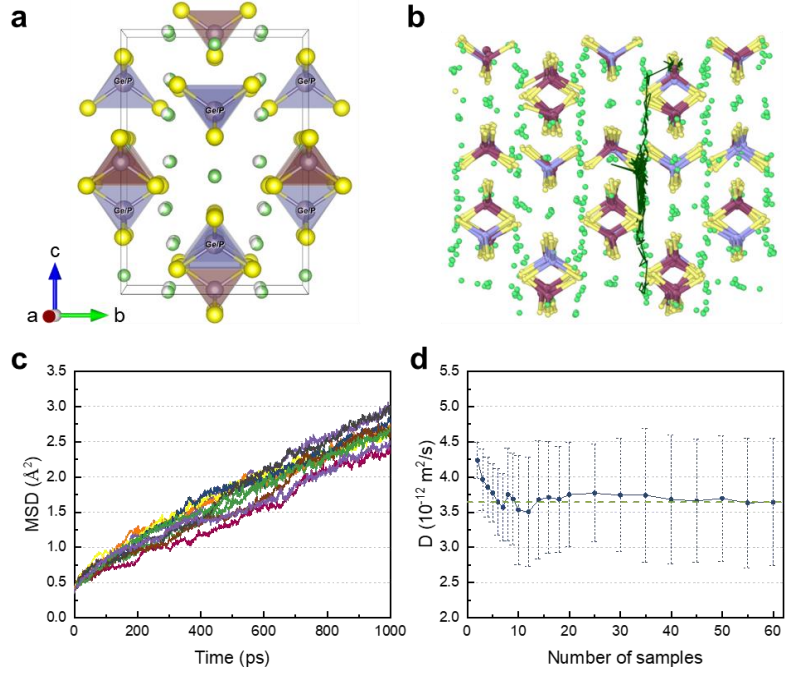

**Figure 1.** The conventional MSD analysis of $Li_{10}GeP_2S_{12}$. (a) Crystal structure of $Li_{10}GeP_2S_{12}$. Yellow atoms: S; small green atoms: fully occupied Li sites; small green-white atoms: partially occupied Li sites; blue tetrahedra: $(Ge_{0.5}P_{0.5})S_4$; and brown tetrahedra: $PS_4$. (b) The diffusion trajectory of a Li atom in 4000 ps. (c) MSD values extracted from 10 individual MD simulations at 300 K over 1000 ps. (d) Averaged diffusion coefficient (*D*) and standard error derived from MSD curves with different number of samples. The dashed green line represents the averaged *D* value of all 60 samples.

We first investigated the Li diffusion in $Li_{10}GeP_2S_{12}$ crystal, one of the most promising superionic conductors due to its exceptionally high ionic conductivity. The crystal has a tetragonal unit cell with the space group $P4_2/nmc$ (137) [6]. As shown in Fig. 1a, there are two tetrahedral sites, 2b (brown) and 4d (blue), in the unit cell. The 4d tetrahedral site can be occupied by both Ge and P atoms, introducing the Ge/P disorder in the crystal. Figure 1b depicts the diffusion trajectory of a Li atom within the $Li_{10}GeP_2S_{12}$ crystal. Over a simulation period of 4000 ps, the Li atom traverses approximately 15 Å along the c-axis but only about 3 Å within the ab-plane, highlighting a pronounced anisotropy in its diffusion behavior [17].

We calculated the conventional MSD values from MD trajectories of $Li_{10}GeP_2S_{12}$ 6×6×4 supercells over 1 ns at 300 K. MSD values can be calculated as follows:

$$\text{MSD}(t) = \frac{1}{N} \sum_{i=1}^{N} [r_i(t) - r_i(0)]^2 \qquad (1)$$

where $t$ is the time for which the MSD is calculated. $N$ is the number of atoms, $r_i(t)$ is the coordinates of the $i^{th}$ atoms at time $t$. Fig. 1c shows the conventional MSD curves for 10 samples, which present significant data fluctuations and low linearity. For conventional MSD calculation, statistical uncertainty originates from the randomness of rare events, i.e., ionic hopping. These issues are particularly critical when the MSD reflects non-Gaussian diffusion behaviors characterized by heavy tails in the ionic displacement probability distribution, rather than the Gaussian distribution expected in normal Brownian motion. For instance, in scenarios involving Lévy flights, rare, large displacements follow a power-law distribution, resulting in a heavy-tailed displacement profile [30]. Lévy flights introduce significant fluctuations into the MSD, potentially distorting the estimated diffusion coefficient if conventional averaging methods are applied. To decrease this uncertainty, a large number of MD simulations are required until the estimated diffusion coefficient value converges.

In this case, we applied 60 individual MD simulations to obtain a converged diffusion coefficient ($D$). For each simulation, we estimated $D$ by performing linear regression on the MSD curve over the time interval from 250 ps to 600 ps. Figure 1d displays the averaged $D$ values and the associated standard error calculated among varying number of samples from 2 to 60. With small number of samples, the averaged $D$ value was highly overestimated and the standard error was underestimated. As the number of samples increases, both the averaged $D$ value and the standard error become more stable. Our results indicate that a large number of samples up to 60 are required to achieve convergence in the $D$ estimate for $Li_{10}GeP_2S_{12}$ at room temperature. However, in practice, the sample capacity for $D$ estimate is often limited and typically no more than 10. With such a small number of samples, the $D$ values derived from conventional MSD curves exhibit significant fluctuations. These results indicate that the conventional MSD fails to capture the main diffusion characteristics, leading to large statistical uncertainty of $D$ estimates with limited samples. Thus, a more reliable estimate of $D$ with limited sample capacity is required in scenarios where computational constraints limit the number of simulations.

**Time-averaged MSD calculation**

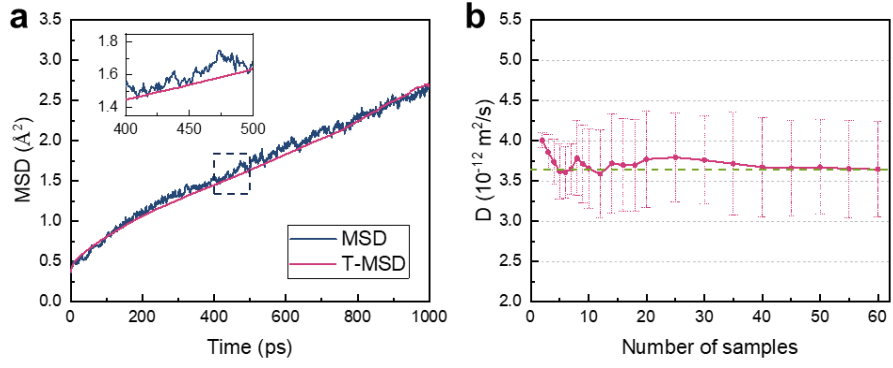

**Figure 2.** Time-averaged MSD curves of $Li_{10}GeP_2S_{12}$. (a) The MSD and time-averaged MSD curves extracted from MD simulation trajectory of 1000 ps. The zoom-in of dashed square. (b) Averaged $D$ and standard error derived from time-averaged MSD curves with different number of samples. The dashed green line represents the averaged $D$ value of all 60 samples.

To identify the reliability of our method, we performed T-MSD analysis on the same $Li_{10}GeP_2S_{12}$ system and compared the results with the conventional MSD analysis. The first part of T-MSD analysis is time-averaged MSD, which quantifies the average squared distance traveled in a certain time interval $\tau$. It can be calculated as follows:

$$\text{MSD}(\tau = m\Delta t) = \frac{1}{M-m}\frac{1}{N}\sum_{t=0}^{M-m}\sum_{i=1}^{N}[r_i(t\Delta t + \tau) - r_i(t\Delta t)]^2 \qquad (2)$$

where $\Delta t$ is the time step in the trajectory, $\tau = m\Delta t$ is the time interval for which the MSD is calculated, and $M\Delta t$ is the total time duration of the trajectory. $N$ is the number of atoms, $r_i(t)$ is the coordinates of the $i^{th}$ atoms at time $t\Delta t$. Time-averaged MSD and conventional MSD curves were extracted from the same MD trajectory of a 6×6×4 supercell of $Li_{10}GeP_2S_{12}$ over 1 ns. As shown in Fig. 2a, these two curves exhibit consistent trends, while the zoomed-in time-averaged MSD curve presents superior linearity and reduced data fluctuations. In MSD analysis, a common practical challenge for accurate $D$ estimation is determining the appropriate time region for linear regression, as deviations or fluctuations outside the linear region can lead to inaccurate $D$ values. The time-averaged MSD method effectively addresses this issue by smoothing out short-term variations and enhancing the linearity of the curve, making it easier to identify the correct region for linear regression.

We then conducted time-averaged MSD calculation on the same 60 MD trajectories we used before. Fig. 2b displays the averaged $D$ values and the associated standard error calculated among varying number of samples from 2 to 60. For comparison, we used the same procedure to estimate $D$ values and standard errors as in Fig. 1d, thus BJ resampling was not employed to determine the standard error. The averaged *D* value derived from time-averaged MSD curves converges to the same value as that obtained from conventional MSD curves, indicating a close agreement between the two methods. It should be pointed out that the *D* values estimated from time-averaged MSD curves converge more rapidly than those obtained from conventional MSD. The *D* value converges when the number of samples exceeds 40, which is only 60% of the samples required by conventional MSD. Moreover, the standard error of *D* derived from time-averaged MSD is $0.592 \times 10^{-12}$ m$^2$/s for the 60 samples, which is 65% of the standard error associated with conventional MSD, indicating an enhanced statistical reliability.

**BJ resampling**

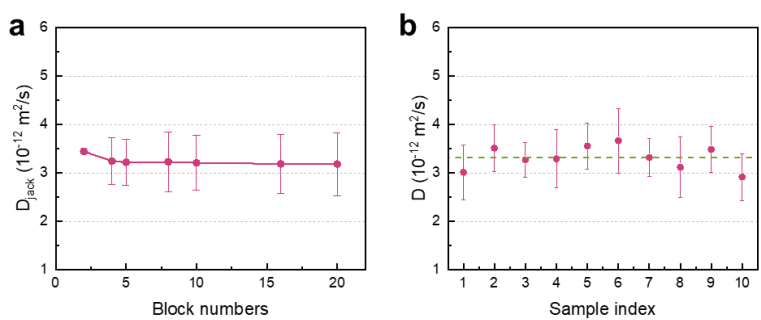

**Figure 3.** BJ resampling of MD trajectories of Li$_{10}$GeP$_2$S$_{12}$. (a) The estimated $D_{\text{jack}}$ converges with increasing block numbers for a MD trajectory over 8 ns. (b) The estimated *D* and confidence intervals from T-MSD analysis compared with the *D* mean value for 10 individual MD simulations.

Then we conducted BJ resampling [31], where the trajectory data is divided into blocks, and different blocks are systematically removed to form subsets. Then the time-averaged MSD analysis is applied to each subset. The BJ resampling allows for the assessment of the sensitivity of the *D* estimate to specific segments of the trajectory, including those that might contain rare events. If a significant change in the *D* estimate occurs when particular blocks are excluded, it indicates that those blocks contain rare events that are skewing the overall estimate. Here we conducted DPMD simulations over a long term up to 8 ns, since the simulation time significantly affects the estimated *D* values as we discussed later (Fig. 4). The estimated $D_{\text{jack}}$ for different block numbers were presented in Fig. 3a. The $D_{\text{jack}}$ converges with increasing block numbers, indicating that the method effectively reduces statistical bias and accounts for time-correlations within the MD trajectories. Furthermore, BJ resampling provides the standard error of *D* estimate, offering a quantitative assessment of its statistical reliability. A smaller standard error value reflects higher accuracy and reliability of *D* estimate, suggesting that the sample *D* is closer to the true *D*. It should be pointed out that BJ resampling allows for standard error calculation from a single simulation by systematically resampling data segments, rather than requiring multiple independent simulations, as is typically necessary in conventional MSD analysis.

Based on standard error value, a confidence interval can be calculated to provide an estimated range for the true $D$. To assess the reliability of this estimate range, we tested whether the confidence interval is consistent with the true $D$ value. Then we performed T-MSD analysis on 10 individual MD simulations of $Li_{10}GeP_2S_{12}$ 6×6×4 supercells over 8 ns. The true $D$ value was determined as the average of the $D$ values from all samples. As shown in Fig. 3, for all 10 samples, the true $D$ value consistently falls within the confidence interval, indicating that the BJ resampling effectively captures the expected range of $D$ values, even in individual samples. Our T-MSD method, therefore, offers a more robust and statistically reliable framework for $D$ estimation than conventional MSD by reducing fluctuations and providing accurate standard error measurements particularly with limited samples.

**Effect of simulation time and sample sizes on $D$ estimation**

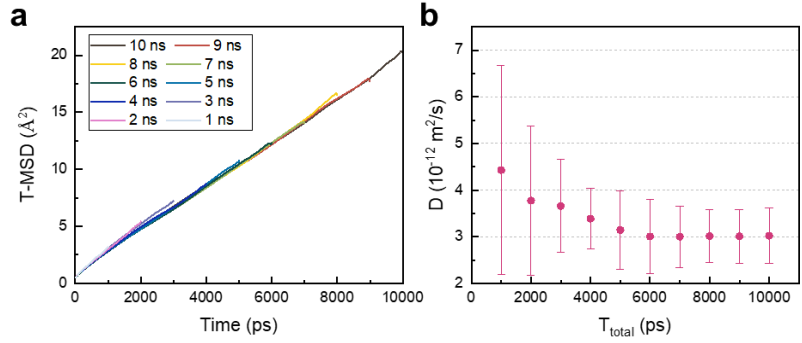

**Figure 4.** Effects of total time on estimates of $D$ by T-MSD analysis of $Li_{10}GeP_2S_{12}$. (a) MSD and time-averaged MSD curves extracted from the same MD simulation trajectory truncated varying time point from 1 ns to 10 ns. (b) The estimated $D$ and standard error vary with total time.

The time scale is critical for the accuracy of $D$ estimate, as longer simulations capture more diffusion events. To assess the effect of simulation time on the $D$ estimates and associated statistical uncertainty, we conducted tests with varying total simulation times. Fig. 4a presents the time-averaged MSD curves extracted from a long-term DPMD simulation at 300 K over 10 ns. These curves were generated by truncating the total simulation at different time points (ranging from 1 ns to 10 ns). All the time-averaged MSD curves exhibit high linearity and consistent trends. Fig. 4b shows the estimated $D$ and standard error from BJ resampling for different total time. The $D$ value for 1 ns is significantly overestimated, suggesting that a shorter simulation time cannot fully capture the diffusion behavior. As the total time increases, the $D$ value converges at around 6 ns, indicating that a sufficient number of diffusion events have been captured by this time, leading to a stable and reliable estimate of $D$. The reduction in standard error over simulation times demonstrates that statistical uncertainty diminishes as the time scale grows, further confirming the accuracy of the $D$ estimate at longer simulation times.

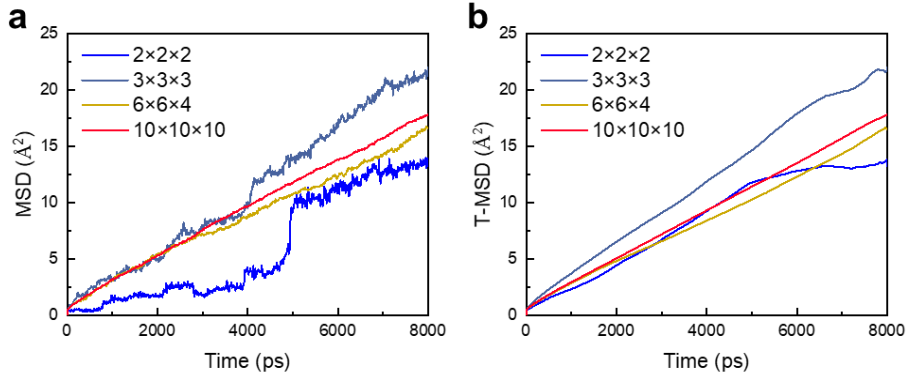

**Figure 5.** Effect of sample size on MSD and T-MSD analysis of $Li_{10}GeP_2S_{12}$. (a) and (b) MSD and time-averaged MSD curves extracted from MD simulations within 2×2×2, 3×3×3, 6×6×4, and 10×10×10 supercells, respectively.

**Table 1.** The estimated $D$ ($10^{-12}$ m$^2$/s), standard error (SE) and coefficient of determination ($R^2$) within different sample sizes.

| Size | $D_{MSD}$ | $R^2_{MSD}$ | $D_{T-MSD}$ | $R^2_{T-MSD}$ | SE |
|---|---|---|---|---|---|
| 2×2×2 | / | / | 4.107 | 0.999 | 2.546 |
| 3×3×3 | 4.551 | 0.920 | 4.497 | 0.999 | 0.886 |
| 6×6×4 | 2.902 | 0.986 | 3.013 | 1.000 | 0.560 |
| 10×10×10 | 3.648 | 0.999 | 3.539 | 1.000 | 0.467 |

Similar to the simulation time, the system size is of significant importance for $D$ estimation. We investigate how MSD and time-averaged MSD depend on the supercell size by carrying MD simulations of $Li_{10}GeP_2S_{12}$ supercells with the sizes of 2×2×2, 3×3×3, 6×6×4, and 10×10×10. Fig. 5a displays the MSD curves for each supercell. The MSD curve for the 2×2×2 supercell exhibits sharp fluctuations and lacks linearity, indicating that, at such small system sizes, the Li jumps are significantly correlated due to periodic boundary conditions. With this poor linearity, it is difficult to give an accurate $D$ estimate. As the supercell size increases, the MSD curves tend to smooth out and exhibit better linearity. Therefore, we omit MSD curve of 2×2×2 supercell and compute $D$ using linear regression of other MSD curves over the interval from 2000 ps to 4800 ps. The estimated $D_{MSD}$ values are summarized in Tab. 1 as well as the coefficient of determination ($R^2$), where a higher $R^2$ (up to 1) value generally indicates a more accurate linear regression fit to the data. The $R^2$ value increases with the system size, indicating that a more reliable estimate of $D$ characteristics in the larger supercells.

For comparison, Fig. 5b displays the time-averaged MSD curves derived from the same MD simulations as those used for the conventional MSD curves. A notable improvement in the linearity of the time-averaged MSD curves is observed, particularly for the smaller 2×2×2 and 3×3×3 supercells. This enhanced linearity allows a more stable and reliable estimation of $D$ by minimizing potential deviations due to short-term fluctuations in the atomic trajectories, a common issue in conventional MSD analysis of small systems. For instance, the estimated $D_{\text{T-MSD}}$ based on the time-averaged MSD of the 2×2×2 supercell is $4.107 \times 10^{-12}$ m$^2$/s, which is consistent with the values obtained for larger supercell sizes. This demonstrates that the time-averaged MSD method effectively captures the diffusion characteristics, even in limited system sizes. Additionally, the $R^2$ values of time-averaged MSD curves are nearly equal to 1, suggesting an efficient linear regression fit. For example, although $R^2_{\text{MSD}}=0.920$ and $R^2_{\text{T-MSD}}=0.999$ for 3×3×3 supercell may appear similar, the first fit leaves 8% of the variance unexplained, whereas the second leaves only 0.1% unexplained, indicating a significantly higher level of precision in the linear regression. This enhancement in linearity and statistical reliability demonstrates that the use of time-averaged MSD curves offers a more refined and accurate approach to determine $D$ under constrained system sizes, where traditional MSD curves may struggle with noise. Furthermore, the statistical precision is indicated by the standard error values obtained through BJ resampling, enabling more reliable assessments of the material's diffusion properties.

**Effect of temperature and material system on *D* estimation**

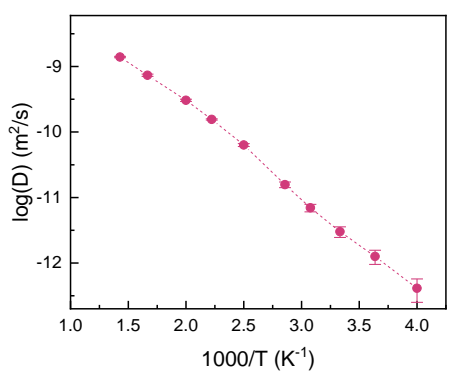

**Figure 6.** The Arrhenius plots of Li diffusion coefficient for Li$_{10}$GeP$_2$S$_{12}$. Data points are obtained at 250, 275, 300, 325, 350, 400, 450, 500, 600, and 700 K.

The diffusion behavior in solid ionic conductors is significantly influenced by temperatures. We also conducted T-MSD analysis for MD simulations from 250 K to 700 K to test the reliability of our method in varying temperature conditions. The increase in temperature significantly enhances diffusion events, promoting a more linear behavior in the MSD curves. As shown in Suppl. Fig. 1, the MSD and time-averaged MSD curves at high temperature exhibit strong consistency, showing similar long-term trends. The estimated $D$ derived from time-averaged MSD curves increased from $4.114 \times 10^{-13}$ m$^2$/s at 250 K to $1.396 \times 10^{-9}$ m$^2$/s at 700 K. Fig. 6 shows the Arrhenius plot for Li$_{10}$GeP$_2$S$_{12}$. It should be noted that the slope changes slightly around 350 K. At high temperatures, the activation energy is 0.248 eV, and it increases to 0.262 eV at low temperatures. This calculated barrier is in remarkable agreement with the experimentally determined barrier of 0.25 eV at room temperature [6]. The standard error estimates obtained using the BJ resampling align with the expected trend that calculations at lower temperatures exhibit greater difficulty and statistical uncertainty compared to higher temperatures. This further supports the conclusion that the proposed BJ resampling effectively provides a reliable estimation of the statistical error in these computations. These results demonstrate that T-MSD analysis is robust and provide reliable $D$ estimates, even under varying temperature conditions.

The diffusion mechanisms of ions can vary significantly across different solid ionic conductors, influenced by factors such as crystal structure, defect chemistry, composition, and the presence of disorder. To evaluate the reliability of the T-MSD method in different materials, we also performed T-MSD analysis for $D$ estimate of Li$_6$PS$_5$Cl crystal with 25% Cl@4c disorder. Suppl. Fig. 2a demonstrates the crystal structure of Li$_6$PS$_5$Cl, where the 4a and 4c sites are randomly occupied by S and Cl atoms. For this analysis, the Cl occupancy on the 4a site was set to 25%. As illustrated in Suppl. Fig. 2b, the conventional MSD curve exhibits noticeable fluctuations at 4000 ps, which can obscure the reliability of the linear regression for $D$ estimate. In contrast, the time-averaged MSD curve maintains its linearity, effectively suppressing short-term fluctuations. Suppl. Fig. 2c shows that the estimated $D_{\text{jack}}$ converges with increasing block numbers, indicating that the method effectively accounts for time-correlations within diffusion events in Li$_6$PS$_5$Cl crystal. This highlights the robustness of the T-MSD method in providing reliable $D$ estimates in different systems.

**Conclusions**

In summary, we proposed an improved post-processing strategy, T-MSD, to address the challenges associated with rare events and auto-correlation in the calculation of diffusion coefficient. T-MSD analysis gives smooth MSD curves by applying time-averaged MSD analysis, ensuring high quality linear regression of $D$ estimates. By applying BJ resampling, T-MSD can provide reliable estimates of standard error directly from a single simulation, whereas traditional methods typically require multiple independent simulations to achieve comparable statistical accuracy.

Using DPMD simulations, we demonstrate that T-MSD effectively reduces statistical uncertainty and enhances the robustness of diffusion coefficient calculations of a typical solid ionic conductor, $Li_{10}GeP_2S_{12}$. Our results also show that T-MSD is versatile, delivering consistent performance across systems of varying sizes and MD simulations with different total durations. The method provides a practical framework for accurately extracting diffusion coefficients while ensuring that proper error estimates are included.

## Methods

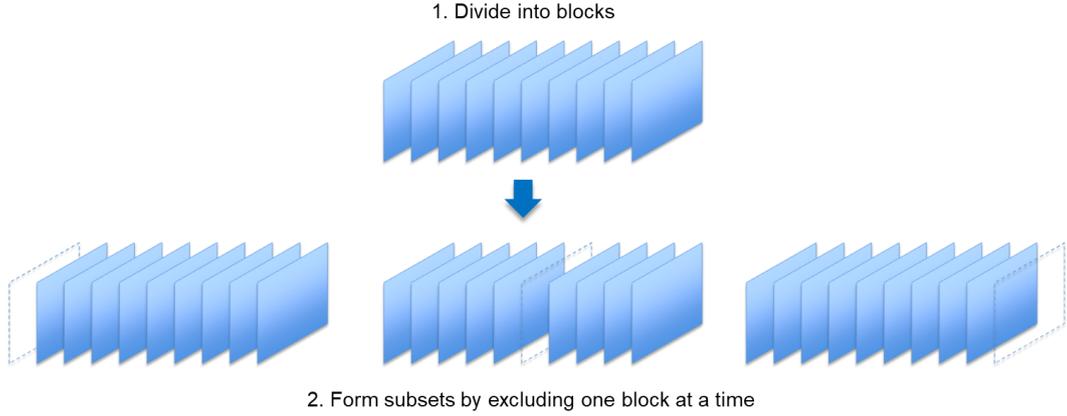

**Figure 7.** Schematic of BJ resampling.

In BJ resampling, the trajectory data is divided into blocks, and different blocks are systematically removed to form subsets, as shown in Fig. 7. First, the original data $X$ is divided into $K$ non-overlapping blocks of size $b$:

$$X = \bigcup_{k=1}^{K} B_k \tag{3}$$

where $B_k = \{x_{(k-1)b+1}, x_{(k-1)b+2}, \ldots, x_{kb}\}$ for $k = 1, 2, \ldots, K$. For each block $B_k$, create a jackknife subset by excluding $B_k$ from $X$:

$$\theta_{(k)} = f(X \setminus B_k) \tag{4}$$

This results in K jackknife subsets $\theta_{(1)}, \theta_{(2)}, \ldots, \theta_{(K)}$. Then one can compute the jackknife estimate of variance and standard error (SE) of $D$:

$$\mathrm{Var}_{\mathrm{BJ}}(\theta) = \frac{K-1}{K} \sum_{k=1}^{K} \left(\theta_{(K)} - \bar{\theta}\right)^2 \tag{5}$$

$$\mathrm{SE}_{\mathrm{BJ}}(\theta) = \sqrt{\mathrm{Var}_{\mathrm{BJ}}(\theta)} \tag{6}$$

## Data availability

All the DP models and configurations of $Li_{10}GeP_2S_{12}$ and $Li_6PS_5Cl$ generated in this study have been deposited in the AIS Square database under accession code: https://www.aissquare.com/models/detail?pageType=models&name=SSE_LGPS_LPSC&id=296. Our training dataset is part of the SSE-abacus datasets deposited in the AIS Square database under accession code: https://www.aissquare.com/datasets/detail?name=SSE-abacus&id=260&pageType=datasets.

## Acknowledgements

This work was supported by the National Key R&D Program of China (No. 2022YFA1403000 and Grants No. 2021YFA0718900), Key Research Program of Frontier Sciences of CAS (Grant No. ZDBS-LY-SLH008), National Nature Science Foundation of China (Grants No. 12374096 and No. 92477114)

## Author Contributions

Z. Z. conceived the project and supervised the research. X. D. Cao designed the analytical method. Y. G. performed the deep potential training and molecular dynamics simulations. Y. G. organized and wrote the paper. X. D. Cao helped to revise the paper and provide scientific discussion when this study encountered problems.

## Competing interests

The authors declare no competing interests.